\documentstyle[preprint,aps,prb,psfig,epsf]{revtex}
\begin{document}
\title{Path integral Monte Carlo simulations of silicates}

\author{Chr. Rickwardt$^{1,}$\footnote{present address:Landesbank Rheinland-Pfalz, Mainz, Germany}, P. Nielaba$^2$, M. H. M\"user$^1$, K. Binder$^1$}

\address{$^1$ Institut f\"ur Physik,
Johannes Gutenberg Universit\"at Mainz, 55099 Mainz, Germany \\ 
$^2$ Fachbereich Physik, Universit\"at Konstanz,
Fach M 691, 78457 Konstanz, Germany \\
}

\date{\today}

\maketitle

\begin{abstract}
We investigate the thermal expansion of crystalline SiO$_2$ in
the $\beta$-- cristobalite and the $\beta$-quartz structure
with  path integral Monte Carlo (PIMC)  techniques.
This simulation method allows to treat
low-temperature quantum effects properly.
At temperatures below the Debye temperature,
thermal properties obtained with PIMC agree better
with experimental results than those obtained with classical Monte
Carlo methods.
\end{abstract}


\section{Introduction}

The study and analysis of materials properties of crystalline 
silicates are important since these systems are used in many
industrial processes and they occur also in many natural rocks.
Many interesting effects have been found by 
experimental techniques at temperatures well below the Debye temperature.
In this temperature range quantum effects like 
zero point motions and corresponding delocalizations of atoms
are important
which have to be taken into account in serious theoretical studies.
Usually this is done by lattice dynamics theories, in the framework of
the harmonic or quasiharmonic approximation. However, both near
second-order structural phase transitions and quite generally
at higher temperatures the accuracy of this approach is sometimes 
uncertain~\cite{rlmb}, 
and methods that work at all conditions would be desirable.

PIMC simulations~\cite{rnb3} enable 
us to analyse the crystal low temperature thermal properties.
In principle, this method yields exact quantum-statistical
averages (apart from statistical errors) and reduces to classical
statistical averages at high temperatures.  
In general the agreement with experimental data is better
compared to classical computations. 
For $\beta$--cristobalite it turns out that even at temperatures as high
 as 600~K only with PIMC a good agreement with experimental findings is
 obtained.
The negative thermal expansion in $\beta$-quartz however may be explained 
by a classical modelling.

The outline of this paper now is as follows: in Sec.~II we define
the models that are studied, in Sec.~III we briefly review the PIMC
technique, while Sec.~IV describes our results in detail. We conclude
in Sec.~V with a brief summary.

\section{The model}

Starting from the diamond lattice of silicon solids
and placing between each neighbour pair of Si atoms an O atom,
we arrive at a silicate which exists in nature in cubic symmetry:
$\beta$- cristobalite.

For the computation of SiO$_2$--structures a variety of potentials
are studied in the literature, for a review see ref.~\cite{rnb10}.
Concerning the computational time it is of great advantage
that apparently two-body potentials like the TTAM--potential~\cite{tsuneyuki88}
and the BKS--potential~\cite{bks} describe the system properties
at least as good as three body potentials. 
Thus most of the investigations presented here have been done with
the TTAM-- and the BKS-- potentials. Comparative studies have
been done as well with other potentials like the two--body contribution
of a complicated potential obtained via ab-initio 
computations~\cite{schroeder96} (``core-shell''-potential). 

In case of the TTAM and the BKS potentials the interactions
between the Si and O-- ions are modelled by pair potentials
of the form:

\begin{equation}
\Phi(r_{ij}) = \frac{Q_i Q_j e^2}{r_{ij}} + A_{ij} \exp(-B_{ij} r_{ij})
- \frac{C_{ij}}{r^6_{ij}}
\label{ttam-formel}
\end{equation}
Here  $e^2 = \frac{1602.19}{4 \cdot 8.8542 \cdot\pi } eV \AA$.
In Eq.(\ref{ttam-formel}) each ion carries a partial charge:
A Si- ion has the charge $Q=2.4$ and an O- ion has the charge $Q=-1.2$.

The TTAM potential has been developed by Tsuneyuki et.al.~\cite{tsuneyuki88}.
The parameters in Eq.(\ref{ttam-formel}) are listed in table~I.
The parameters in  Eq.(\ref{ttam-formel}) for the BKS- potential~\cite{bks} 
are tabulated in table~II.

In general complex materials with several components
interacting with
long ranged Coulomb interactions have to be handled by
appropriate numerical procedures such as the Ewald summation
technique. In the numerical simulation the computation of the 
Coulomb contribution to the energy requires a summation 
not only over the particle indices but also 
over all periodic boxes:
\begin{equation}
E_{coul} = \frac{1}{2} \sum_{\vec{n}} \sum_{{i, j} \atop { \ (i \not= j \ {\rm for}\  \vec{n} = \vec{0})}}
\frac{Q_i Q_j e^2}{|\vec{r}_{ij} + \vec{n}|}
\label{coulombenergie}
\end{equation}
The sum in Eq.(\ref{coulombenergie}) is over all periodic
images shifted by $\vec{n}$ and all particle indices except
$i=j$ for $\vec{n} = \vec{0}$.
For numerical details of the computation of the Coulomb energy and how to use
a relatively simple and efficient algorithm we refer to Ref.~\cite{frenkel96}

Both potentials have in common that at small interparticle
distances the atomic repulsion is neglected and rather a divergency in
the potentials at small distances appears. In order to avoid
a ``Coulomb-collapse'' of the system, at small distances
we have to include an additional repulsive term.
This term can be modelled as a harmonic contribution~\cite{kathrin} 
suitable for molecular dynamics simulations or as
a purely hard sphere repulsion. In this study for the sake of
simplicity we
follow the latter prescription, the hard sphere distance is chosen
as the distance of the first potential maximum in Eq.~(\ref{ttam-formel}).
\section{Path integral Monte Carlo simulations}

Canonical averages $<A>$ of an observable $A$ in a system
defined by the Hamiltonian ${\cal H} = E_{kin} + V_{pot}$ 
of $N$ particles in a volume $V$
are given by:
\begin{equation}
<A> = Z^{-1} \quad Sp \quad [ A exp(-\beta{\cal H})] \quad .
\end{equation}
Here $Z = Sp \quad [ exp(-\beta{\cal H})]$ is the partition function  and
$\beta = 1/k_B T$ is the inverse temperature.
Utilising the Trotter--product formula,
\begin{equation}
exp(\beta{\cal H}) = \lim\limits_{P\to\infty}
(exp(-\beta{E_{kin}}/P) exp(-\beta{V_{pot}}/P))^P \quad ,
\end{equation}
we obtain the path integral expression~\cite{rnb3}
 for the partition function:
\begin{equation}
     Z(N,V,T) 
        = \lim\limits_{P\to\infty}
	(\frac{mP}{2 \pi \beta \hbar^2})^{3NP/2}
	\prod\limits_{s=1}^P
	\int d\{{\bf r}^{(s)}\}  
    {\exp}[-\frac{\beta}{P} (
   {\displaystyle \sum\limits_{k=1}^N} \frac{mP^2}{2\hbar^2\beta^2}
	    ({\bf r}_k^{(s)} - {\bf r}_k^{(s+1)})^2 
	    +V_{pot}(\{{\bf r}^{(s)}\}))] 
\label{Z}
\end{equation}
Here, $m$ is the particle mass, integer $P$ is the Trotter number
and ${\bf r}_k^{(s)}$ denotes the coordinate of particle $k$
at Trotter-index $s$, and periodic boundary conditions apply,
the particle with Trotter-index $P+1$ is the same as the 
particle with Trotter-index $1$. 
This formulation of the partition function
allows us to perform Monte Carlo simulations for increasing values
of $P$ approaching  the true quantum limit for $P \to \infty$.
Note that Eq.~(\ref{Z}) does not take into account any 
quantum- mechanical exchange between particles (for atoms as heavy
as Si and O this is an excellent approximation, though it would
not work for He crystals).

Thermal averages in the ensemble with constant pressure $p$
are given via the corresponding partition function
\begin{equation}
\Delta(N,p,T) = \int_0^\infty dV \exp [-\beta p V] Z(N,V,T)
\end{equation}.
In order to make our results comparable with those of experiments,
our PIMC simulations have been performed in the constant pressure ensemble.
Most of our simulations~\cite{rnb1} have been done with $N=684$ particles
($\beta$--cristobalite), $N=576$ ($\beta$--quartz), $p=0$, 
and $P$-- values up to 100. A typical PIMC data point in 
figure~\ref{e.bcrist}
required about 1200 CPU hours on a  CRAY--T3E (single processor).

\section{Results}

\subsection{Finite size effects and pair potential sensitivity}

\begin{figure}
\begin{picture}(100,100)%
\put(0,0){\psfig{figure=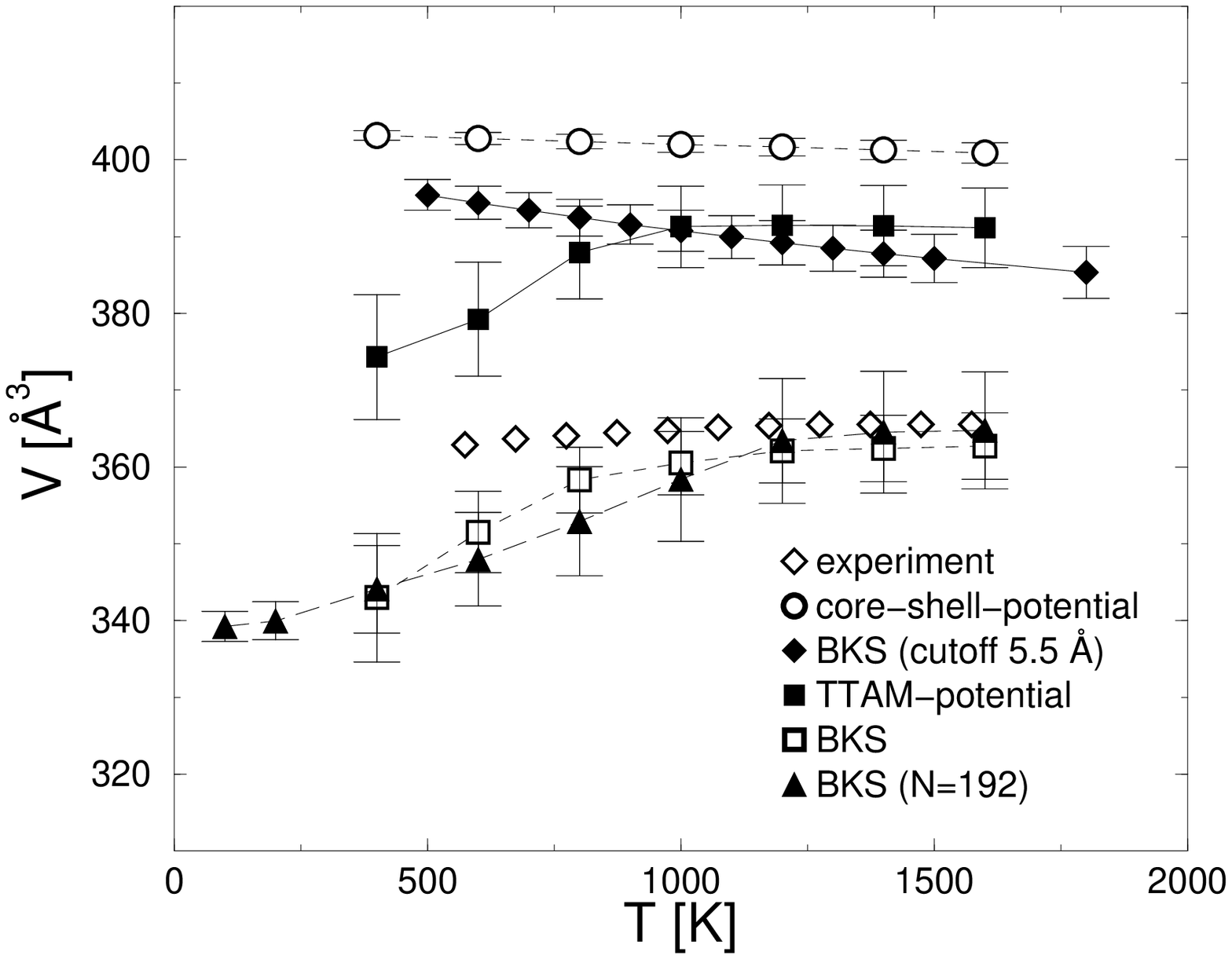,width=100mm,height=100mm}}
\end{picture}
\caption[]{
Classical (P=1) unit cell volume for
$\beta$-cristobalite. The simulation box is a cube containing
27 unit cells with $N=648$ particles, except otherwise noted.
Comparison with experimental data taken from Ref.~\cite{rnb8}.
}
\label{class.bcrist}
\end{figure}

In order to analyse effects due to the finite system size
simulations for the system in the $\beta$-- cristobalite structure
have been done using the three potentials described above and
$N=648$ particles as well as with the BKS-- potential and $N=192$ particles.
In Figs.~\ref{class.bcrist} and \ref{e.bcrist}
the potential energy and the systems box volume
is shown.

The energy  obtained by the core-shell potential (in the two-body reduced form)
is significantly smaller than the energy obtained by the other two
potentials. The volume obtained by the core-shell potential 
is much larger than the experimental data~\cite{rnb8}
and the thermal expansion coefficient is negative over the entire
temperature range studied- in contrast to the experimental results
for $\beta$-cristobalite.

For the BKS-- potentials
the successful description of SiO$_2$-melt properties~\cite{kathrin}
using a  cut-off distance of 5.5$\AA$ for the short ranged interactions 
suggested the choice of this cutoff- distance in the present 
study as well. From figure~\ref{class.bcrist} however it is apparent
that the volume decreases strongly with increasing temperature- in contrast
to the experimental data.
For the ordered crystal it turned out that a larger cutoff value
(12 $\AA$) is more adequate for the description of the system properties.

The energies obtained by the TTAM- potential and by the BKS- potential
(for both system sizes) agree well. In case of the volume however
the results for the BKS- potential agree for both system sizes
within numerical scatter and show a qualitatively similar temperature
dependence as the volume obtained by the TTAM- potential. The quantitative
values of the volume for the TTAM- potential however are much larger than
the experimental data, whereas the volume obtained by the BKS- potential
agrees well with the experimental data for temperatures above 700~K.

In summary the best agreement with the experimental data
can be obtained with the BKS- potential - at least in a temperature 
range above 700~K.
For smaller temperatures apparently none of the classical 
computations can successfully describe the experimental data.

\subsection{Quantum simulations for $\beta$- cristobalite}

\subsubsection{Potential energy and volume}

\begin{figure}
\begin{center}
\begin{picture}(100,100)
\put(0,0){\psfig{figure=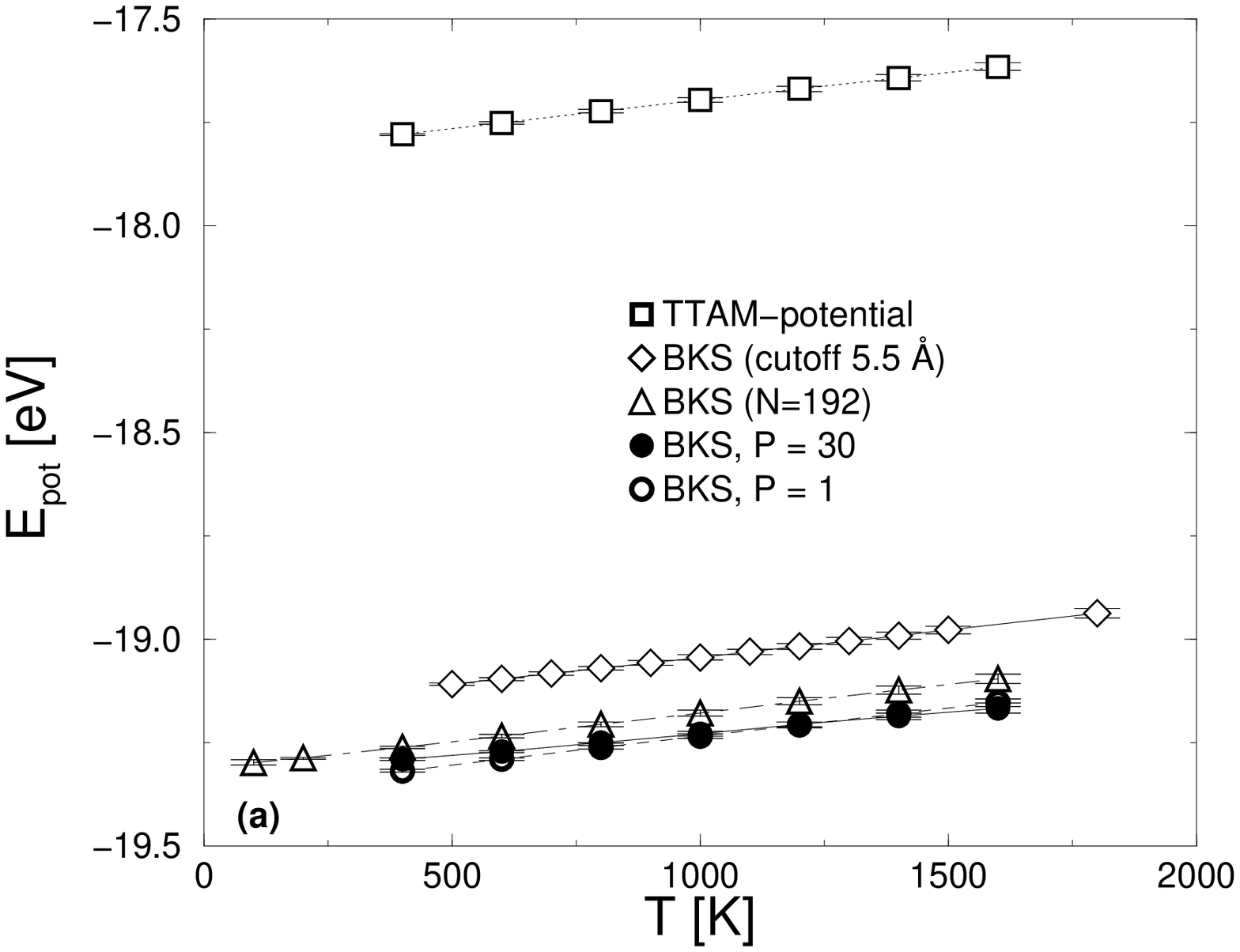,width=100mm,height=100mm} }
\end{picture}
\begin{picture}(100,100)
\put(0,0){\psfig{figure=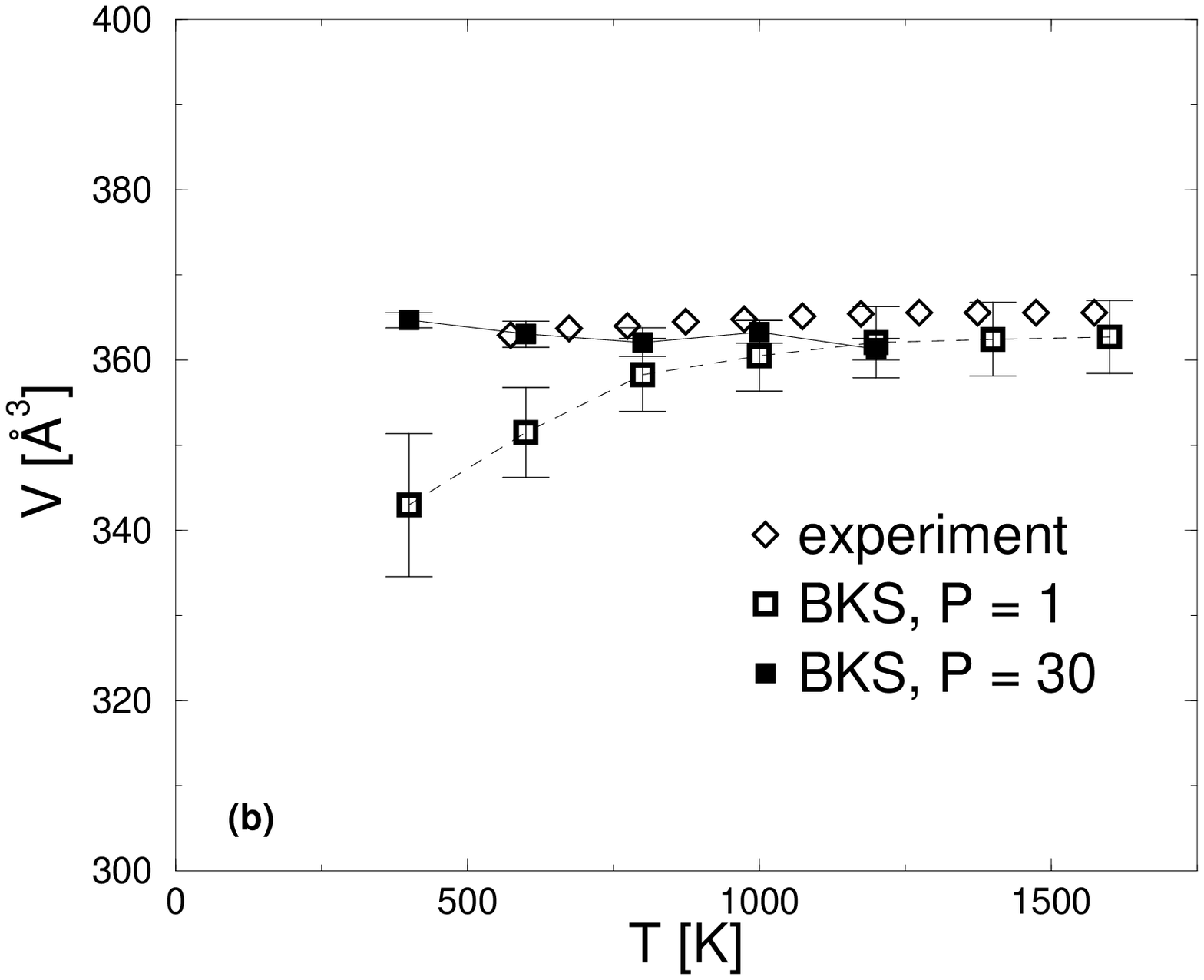,width=100mm,height=100mm}}
\end{picture}
\caption[]{
Potential energy (a) and unit cell volume (b)
for $\beta$-cristobalite, results of path integral Monte Carlo
simulations (with $N=648$ particles and 27 unit cells
unless otherwise noted). 
Comparison with experimental data taken from Ref.~\cite{rnb8}.
In addition simulations have been done with a
``core-shell'' potential~\cite{schroeder96} (see text)
with an average potential energy of about -43.2 eV. 
}
\label{e.bcrist}
\end{center}
\end{figure}

A different picture emerges if results of quantum simulations 
are considered in addition (BKS- potential, $N=648$), see 
figure~\ref{e.bcrist}.
At temperatures below 700~K the quantum results for the potential
energies deviate from the classical
results to higher values, but for higher temperatures they 
agree with the classical values.
The volume of the simulation box obtained by the PIMC simulations
(Fig.~\ref{e.bcrist})
agree within numerical scatter with the experimental data 
for all temperatures studied- in contrast to the classical simulations. 
Despite the apparently high temperatures of 700~K quantum effects
are very important for the thermal properties of $\beta$-cristobalite
and thus quantum effects should not be neglected in simulations
of real materials. This supports the assumption that quantum effects
may still play an important role above the Debye temperature
which is at about 500~K for SiO$_2$.

\subsubsection{Structural details}

According to a model of Wyckoff\cite{wyckoffa,wyckoffb} for
$\beta$- cristobalite the O- atoms are located in the middle of 
a bond connecting two Si- atoms, the Si--O--Si bond angle
is 180$^o$ and the Si-O bond length 1.54~$\AA$.
In order to correct these values to obtain the actually observed
angle (between 140$^o$ and 150$^o$) and bond lengths (Si-O: about
1.61 $\AA$) several models have been suggested in which the 
O- positions deviate from the ideal middle 
places~\cite{wright75,hatch91,dove92a,dove92b,giddy93}.
The question which of these models describes the properties
of $\beta$- cristobalite best is not yet decided by experimental methods.

\begin{figure}
\begin{center}
\begin{picture}(80,80)
\put(0,0){\psfig{figure=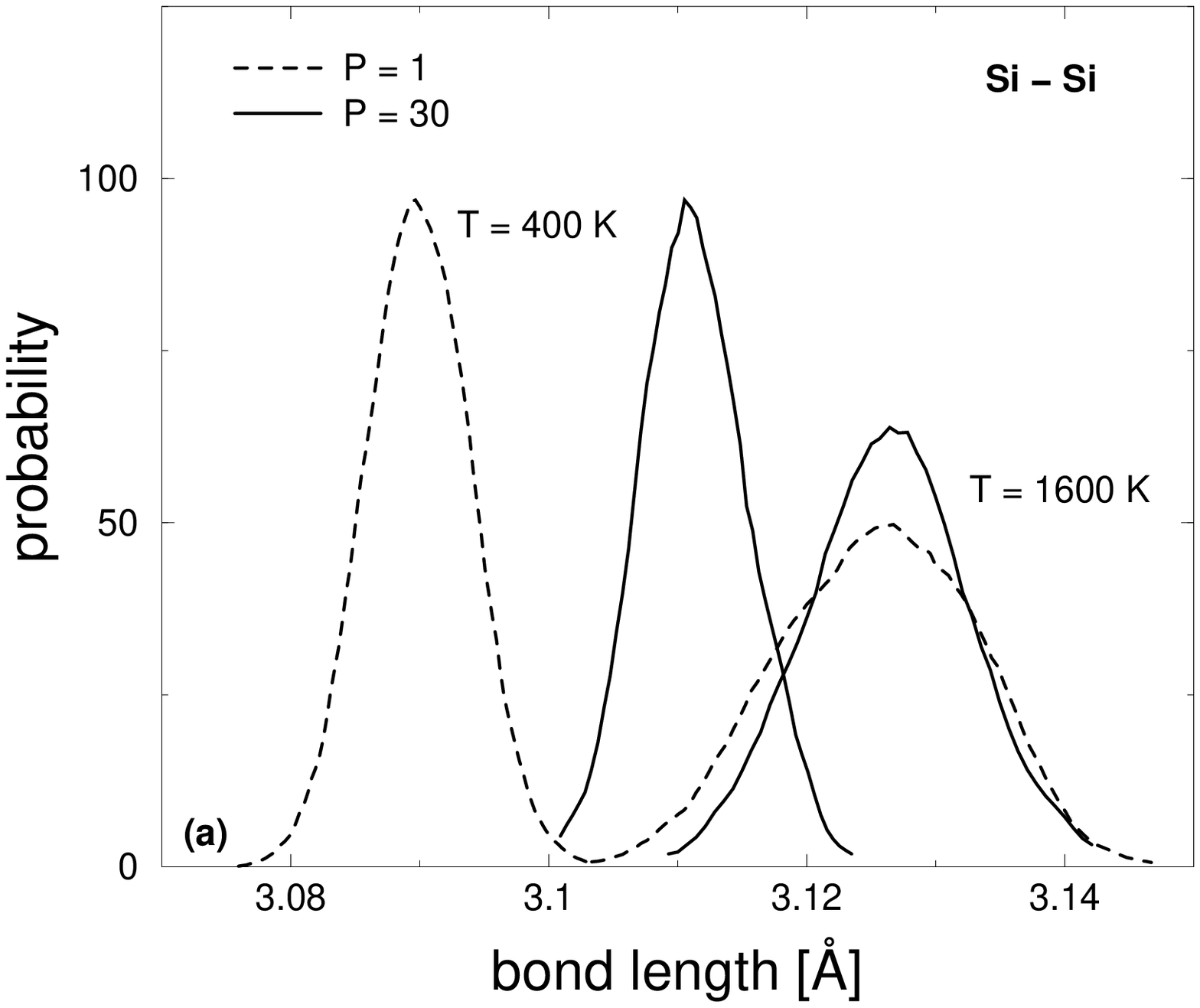,width=80mm,height=80mm}}
\end{picture}
\begin{picture}(80,80)
\put(0,0){\psfig{figure=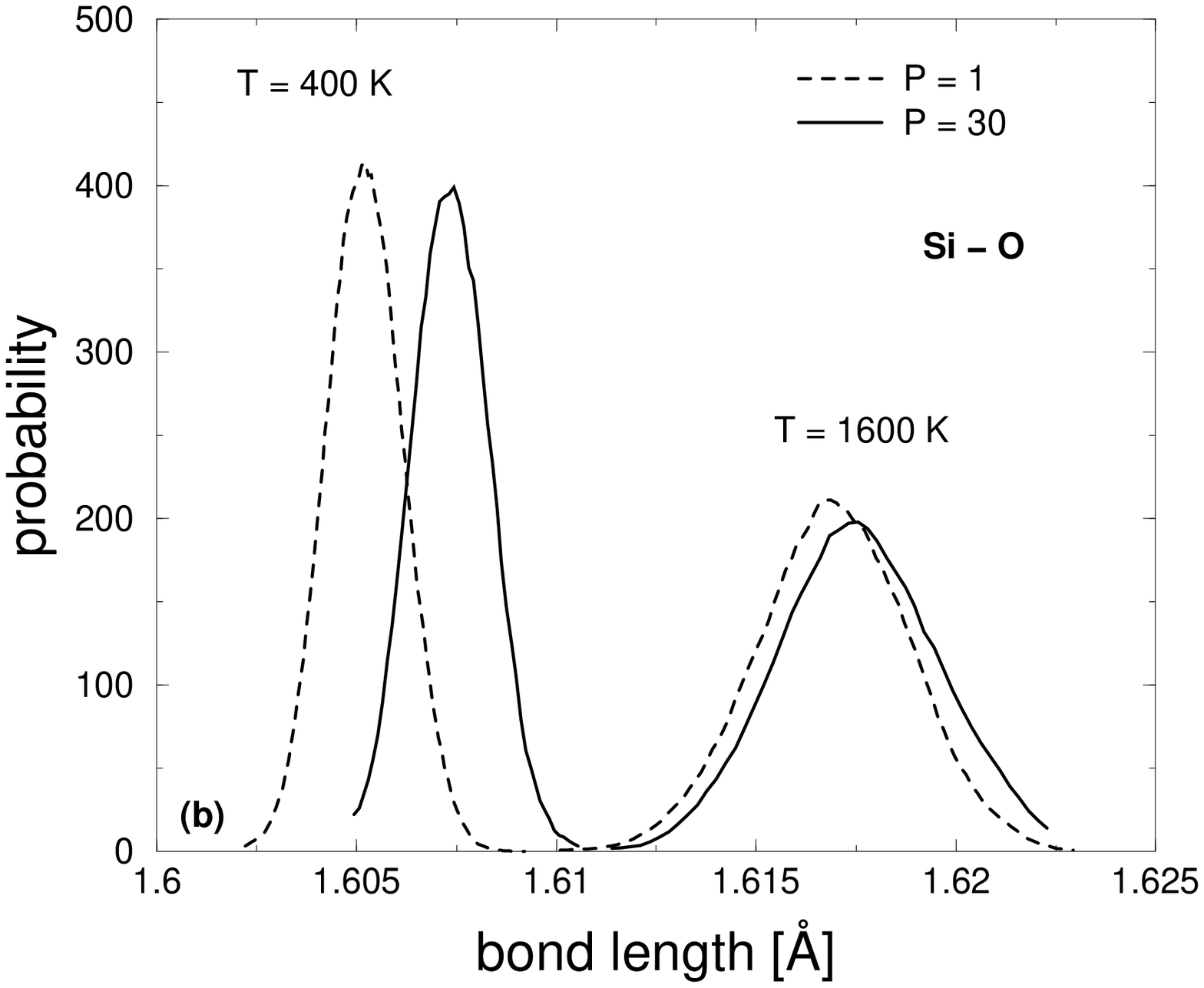,width=80mm,height=80mm}}
\end{picture}
\begin{picture}(80,80)
\put(0,0){\psfig{figure=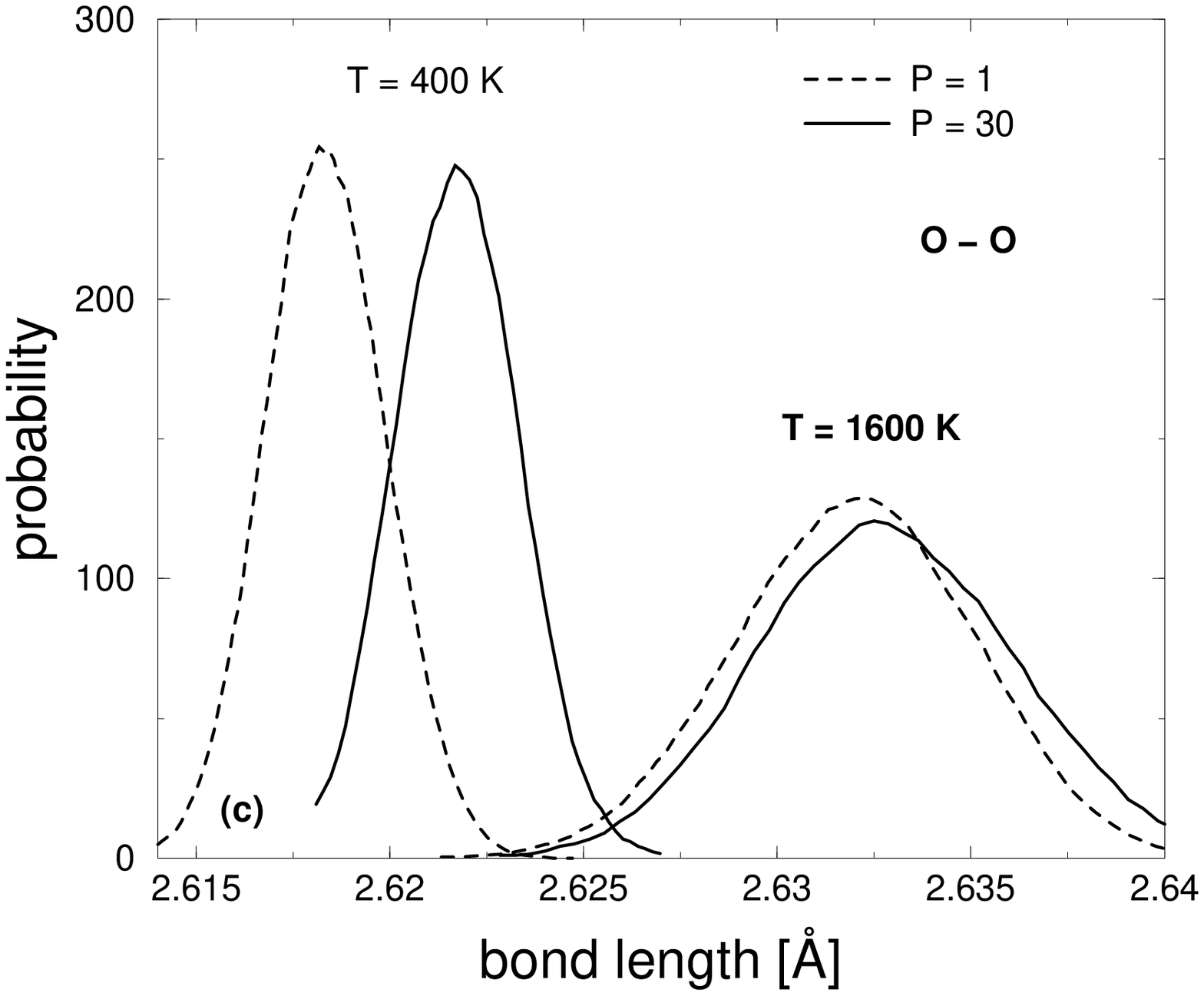,width=80mm,height=80mm}}
\end{picture}
\caption[]
{
Distributions of the bond lengths in $\beta$- cristobalite
at the temperatures T = 400 K and 1600 K. 
(a) Si-Si bonds, (b) Si-O bonds, (c) O-O bonds.
Comparison of classical (P=1) and quantum simulations (P=30).
}
\label{dbonds-bcrist}
\end{center}
\end{figure}

\begin{figure}
\begin{center}
\begin{picture}(80,80)
\put(0,0){\psfig{figure=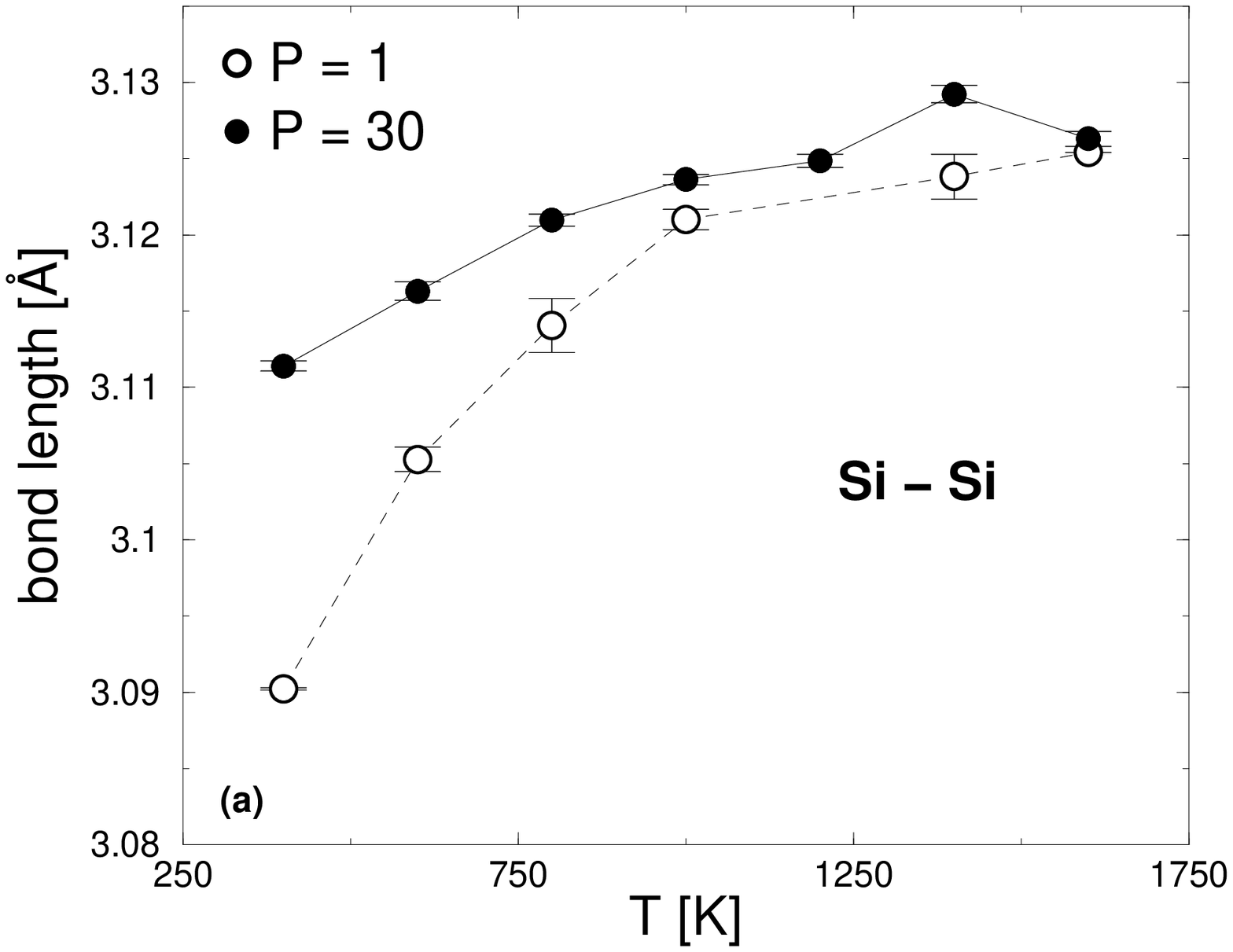,width=80mm,height=80mm}}
\end{picture}
\begin{picture}(80,80)
\put(0,0){\psfig{figure=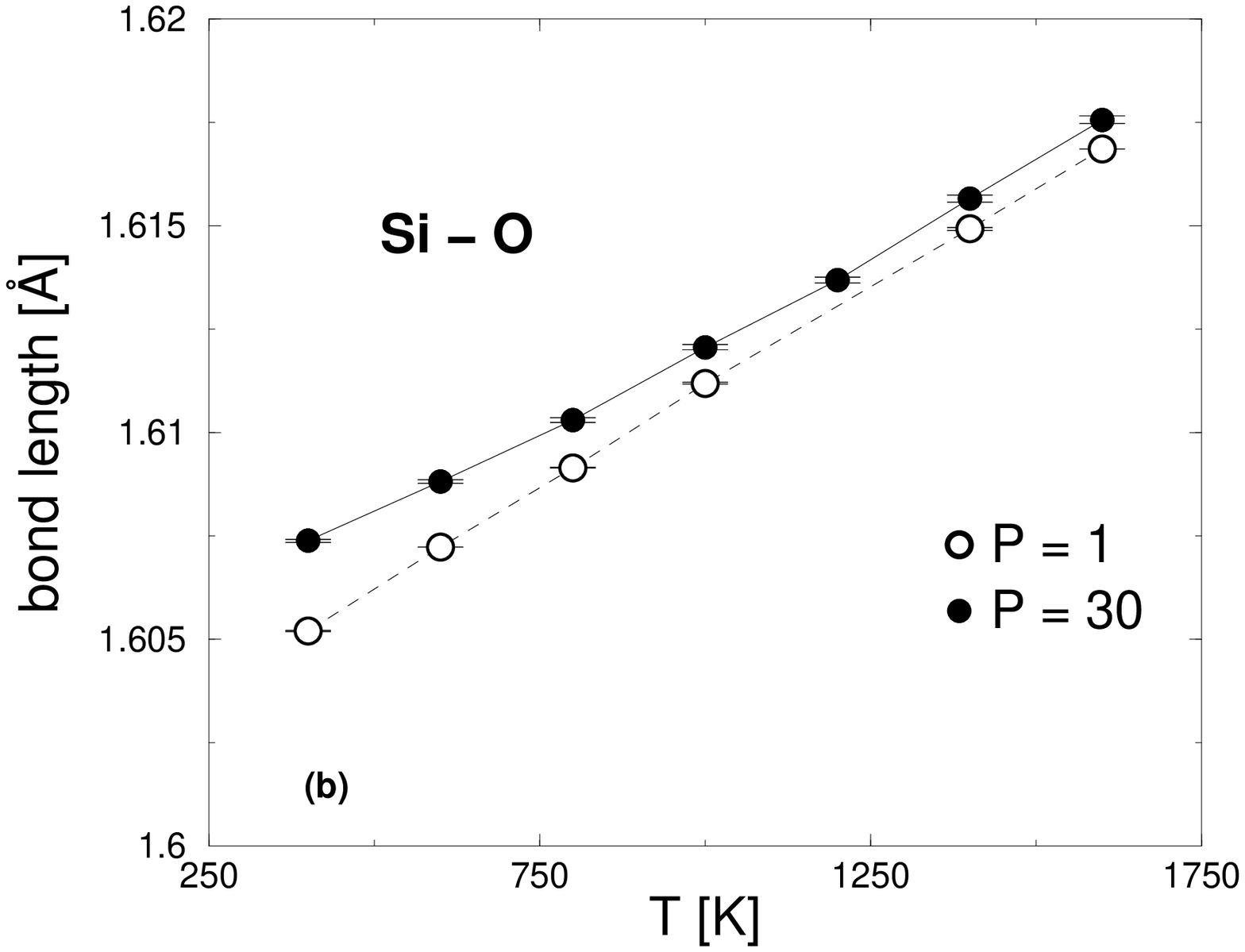,width=80mm,height=80mm}}
\end{picture}
\begin{picture}(80,80)
\put(0,0){\psfig{figure=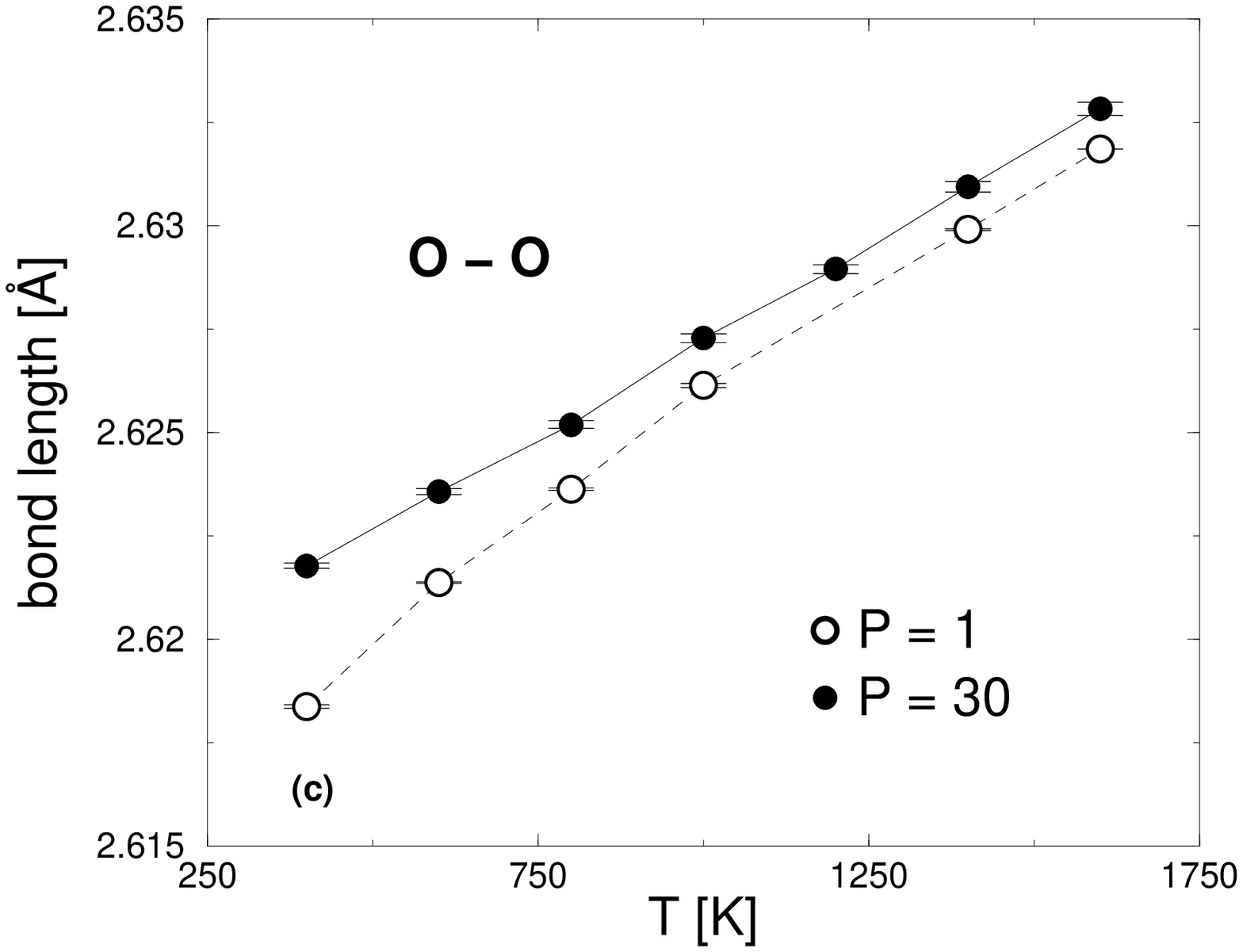,width=80mm,height=80mm}}
\end{picture}
\caption[]
{
Bond lengths between nearest neighbours in $\beta$- cristobalite.
(a) Si-Si bonds, (b) Si-O bonds, (c) O-O bonds.
Comparison of classical (P=1) and quantum simulations (P=30).
}
\label{bonds-bcrist}
\end{center}
\end{figure}

In figure~\ref{dbonds-bcrist} we show the bond length distribution function
for the different atom pairs for three temperatures using the BKS potential
and in figure~\ref{bonds-bcrist} the average bond lengths as a function
of temperature.
Note that due to the choice of a finite size cubic simulation box
with periodic boundary conditions, $\beta$-cristobalite is stable
(or metastable, respectively) in the simulation over the entire 
temperature range, unlike experiment.

In the PIMC simulations (with P=30) the bond lengths have larger
values at low temperatures compared to the classical case 
(P=1) (Fig.~\ref{bonds-bcrist}). The quantum results for the Si-O bond length
are in much closer agreement with the literature value of 1.61~$\AA$.
The Si-O as well as the O-O bond lengths scale approximately linearly
with the temperature, the Si-Si bond length decrease is stronger
in parallel to the temperature dependence of the volume.
This may be due to the fact that the positions of the Si-atoms
fix the lattice constant of $\beta$- cristobalite and the
O- atoms have a much larger mobility in the crystal lattice.
In figure~\ref{bonds-bcrist} we find again the different behaviour 
of the Si-Si bonds compared to the other bonds. The temperature
dependence of the Si-O and the O-O bonds have the following features:
\begin{itemize}
\item The width of the distributions increases with temperature, the height
decreases.
\item The average values increase with the temperature.
\item At the same temperature the distributions obtained by PIMC
 (P=30) have larger average values than the classical distributions.
\item The difference in the distributions and the average 
values between the classical and the quantum results decrease with the
temperature.
\end{itemize}

All these effects are present in the Si-Si bonds as well,
but there is a significant difference:
In Figs.~\ref{dbonds-bcrist} and \ref{bonds-bcrist} we find that
for the other two bond types the differences between the
classical and the quantum values at the temperature $T$ are always smaller
 than the difference between the classical values at $T$ and the next
higher temperature studied. For the Si-Si bond the quantum distributions
at 400~K are closer to the distributions at 1600~K (which do not
differ very much any more) than to the classical distributions at 400~K.
This shows that in particular the Si-Si bond properties
are significantly determined by quantum effects.

\begin{figure}[hbtp]
\begin{picture}(000,150)
\put(0,-20){\epsfxsize=150mm\epsfbox{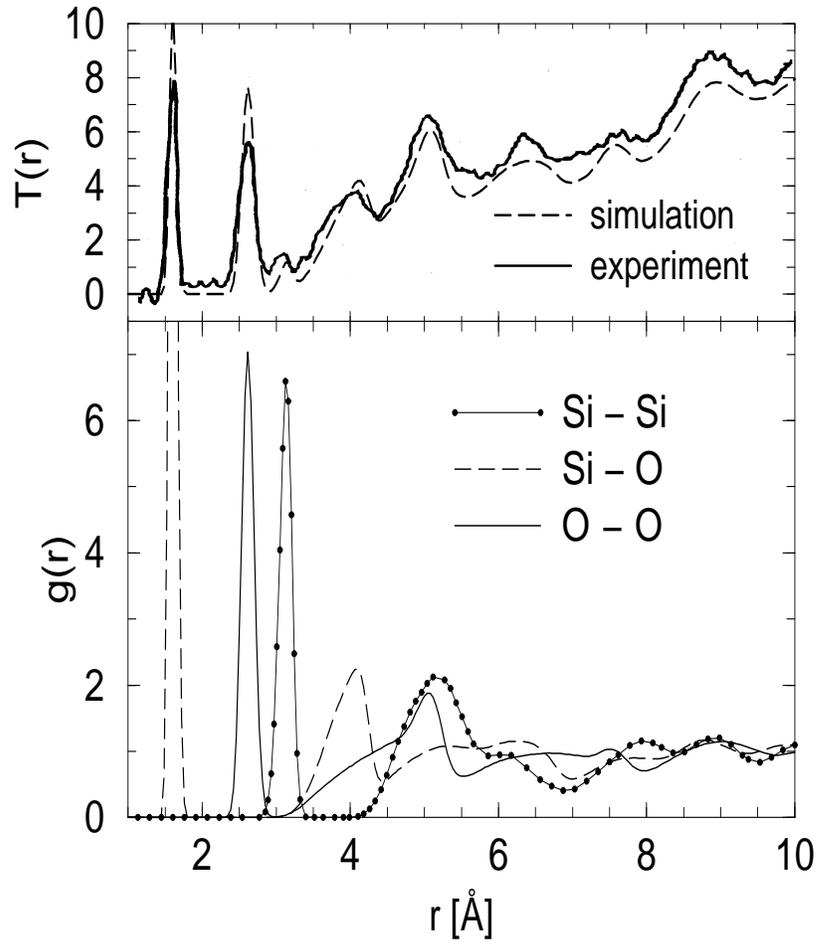}}
\end{picture}
\caption{a) $T(r)$ for $\beta$-cristobalite
as a function of distance $r$ at 573~K.
Experiment is represented by a solid line and simulation
data is represented by a dashed line.
b) Corresponding radial distribution function $g(r)$ for Si-Si, Si-O, and O-O
bonds. The curves are normalized such that $g(r) \to 1$ for $r\to\infty$.
}
\label{martin_crist}
\end{figure}

In order to obtain more information on the long-range molecular structure,
the radial distribution functions $g(r)$ were computed up to a distance
of 10~$\AA$ for Si-Si, Si-O, and O-O bonds. The knowledge of $g(r)$
allows the calculation of the so-called total correlation function
$T(r)$, which is a superposition of the various $g(r)$'s.~\cite{dove97}
$T(r)$ has been measured recently by neutron diffraction.~\cite{dove97}
The measurements allowed to accurately estimate the real Si-O bond lengths
and it was concluded that the typical Si-O bond length is
considerably longer than the Si-O bond length in the so-called ideal
$\beta$-cristobalite geometry. In Fig.~(\ref{martin_crist}a), a comparison
between the experimentally observed and the calculated function $T(r)$ is
shown. The agreement between the two curves is at least semi quantitative
and the conclusions drawn in Ref.~\cite{dove97} could be confirmed with
the exception of the width of the first Si-O peak and the first O-O peak.
These peaks are broadened in experiment with respect to  simulations,
which include quantum effects. The broadenings might be due to lattice
defects or to internal surfaces in the experimental samples.
Both kind of defects were absent in our simulations.
Similar results on $T(r)$ were reported previously by
purely classical simulations~\cite{swainson95b}
on the basis of the TTAM potential.

Fig.~(\ref{martin_crist}b) allows to relate the features in $T(r)$ to
the various $g(r)$'s. It is interesting to note that the peak in
$T(r)$ at about $5\,\AA$ is due to a simultaneous maximum in
$g_{\rm O\,O}(r)$ and $g_{\rm Si\,O}(r)$, while the peak in $T(r)$ at
$6.25$~$\AA$ does not have a corresponding peak in any $g(r)$.
The large local maximum in $T(r)$ at about $9\,\AA$ is then located at a
position where all $g(r)$'s have a local maximum as well. The shape of
$T(r)$ is particularly sensitive to details of the potentials at large
distances, e.g., cutting off the short-range part of the potential at
$4.5\,\AA$ alters $T(r)$ significantly for $r \ge 8\,\AA$.

\subsection{Quantum simulations for $\beta$-quartz}

The $\beta$- quartz structure differs from the $\beta$- cristobalite structure
by the positions of the atoms in the unit cell.
$\beta$- cristobalite has a cubic structure, $\beta$- quartz 
has a hexagonal structure and at a temperatures below 846~K
a trigonal $\alpha$- quartz is
stable~\cite{tse92a,tse92b,tse91,liu93,axe70,kihara90}. 
For structural details of the $\beta$- quartz see appendix~A.
The piezoelectricity (and pyroelectricity) of $\beta$-quartz
is at least reduced for $\beta$- cristobalite due to symmetry reasons.

As for $\beta$- cristobalite the simulations have been done
for Trotter values of P=1 and P=30 at temperatures between
600~K and 1600~K. 
Generalizing Eqs.~(3),(5) and (6),
in our constant pressure simulations a hydrostatic pressure $p$
has been applied and fluctuation trials of the box lengths in $x$-,
$y$- and $z$- directions have been attempted independently
keeping the angles between the box axes constant. 
Lattice constants ($a$ and $c$) have been obtained from the resulting
average box geometry.
The experimentally observed transition at 846~K
is suppressed in the simulations since we keep our box in
rectangular shape. A trigonal $\alpha$- quartz structure thus cannot
be stabilised, rather a metastable, undercooled $\beta$-quartz
is present in the simulations. Above the phase transition temperature,
the structure in the simulation agrees with the structure deduced
from the experimental data. 

\begin{figure}
\begin{center}
\begin{picture}(100,100)
\put(0,0){\psfig{figure=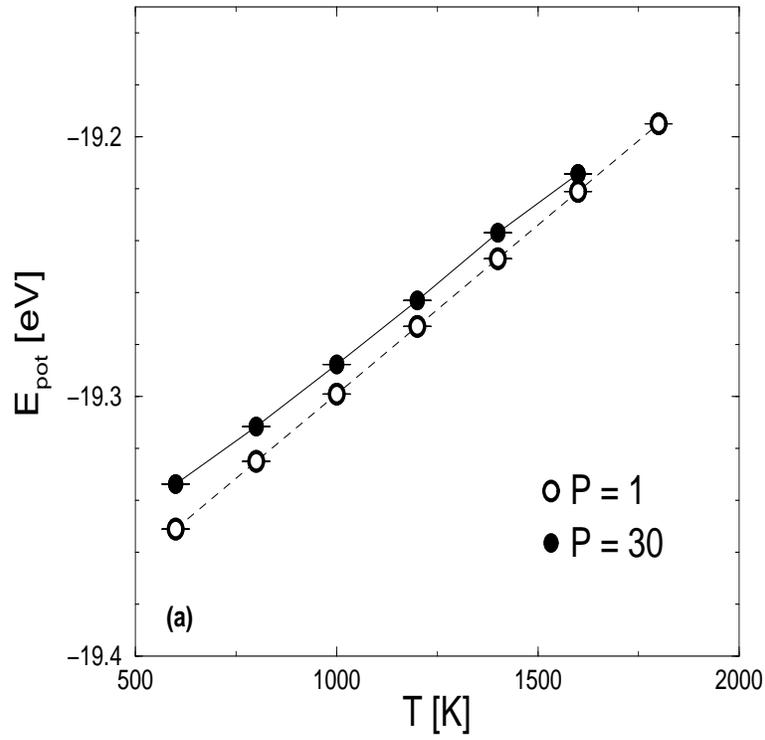,width=100mm,height=100mm}}
\end{picture}
\begin{picture}(100,100)
\put(0,0){\psfig{figure=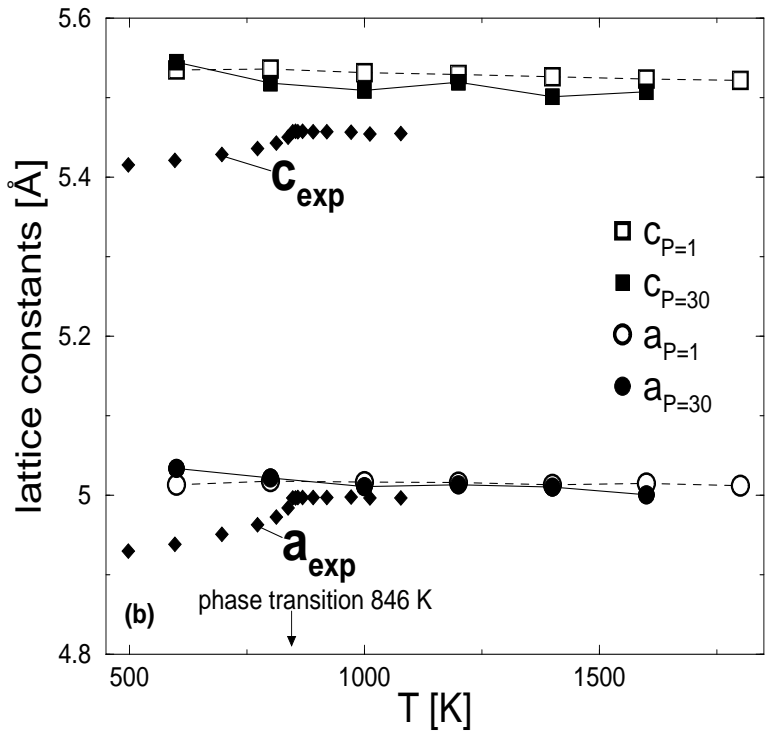,width=100mm,height=100mm}}
\end{picture}
\caption[]{
Potential energy (a) and lattice constants (b) of $\beta$-quartz
using the BKS potential}
\label{bquarz1}
\end{center}
\end{figure}

In figure~\ref{bquarz1} the potential energy and lattice constants
of $\beta$-quartz
are compared to experimental data~\cite{kihara90}.
The energy has a linear temperature dependence in the classical
case and slightly higher values in the quantum simulations,
the difference to the classical values increases the lower the
temperature. Apparently
in this temperature region quantum effects become important-
as can be expected due to a Debye- temperature of 500~K
for SiO$_2$. 
The temperature dependence of the  volume only qualitatively agrees 
with the experimental data. In figure~\ref{bquarz1} we note the
qualitatively different behaviour between the experiments and the
simulation data at temperatures close
to the $ \alpha \leftrightarrow \beta $ transition at 846~K.
As explained above this different behaviour is caused by the
box shape constraint in the simulation.

As in the experiment the lattice constant in a- direction is
approximately temperature independent at high temperatures 
whereas in c- direction a negative thermal expansion coefficient 
is found. In the case of the lattice constant $c$,
 the PIMC results are in somewhat better agreement with
the experimental data than the classical results. 
Since for $\beta$-cristobalite it turned out that by using the literature
value (5.5~$\AA$) for the cutoff the agreement with the 
experimental data is not as good as without the cutoff (see above),
we chose the maximum cutoff value which is possible in
our box geometry.
The deviation of 1\% (c-direction) and 0.3\% (a- direction)
from the experimental values may be due to the choice of 
parameters in the interaction potential, in particular to the
cutoff of the short ranged interactions in the BKS- potential. 
Probably a restriction to smaller cutoff values would 
result in a better agreement with experimental data.

A typical simulation data point in figure~\ref{bquarz1} also 
required a computational effort of about 1200~CPU hours
on a CRAY-T3E computer (in units of single-processor time).
 
\begin{figure}
\begin{center}
\begin{picture}(150,100) 
\put(0,0){\psfig{figure=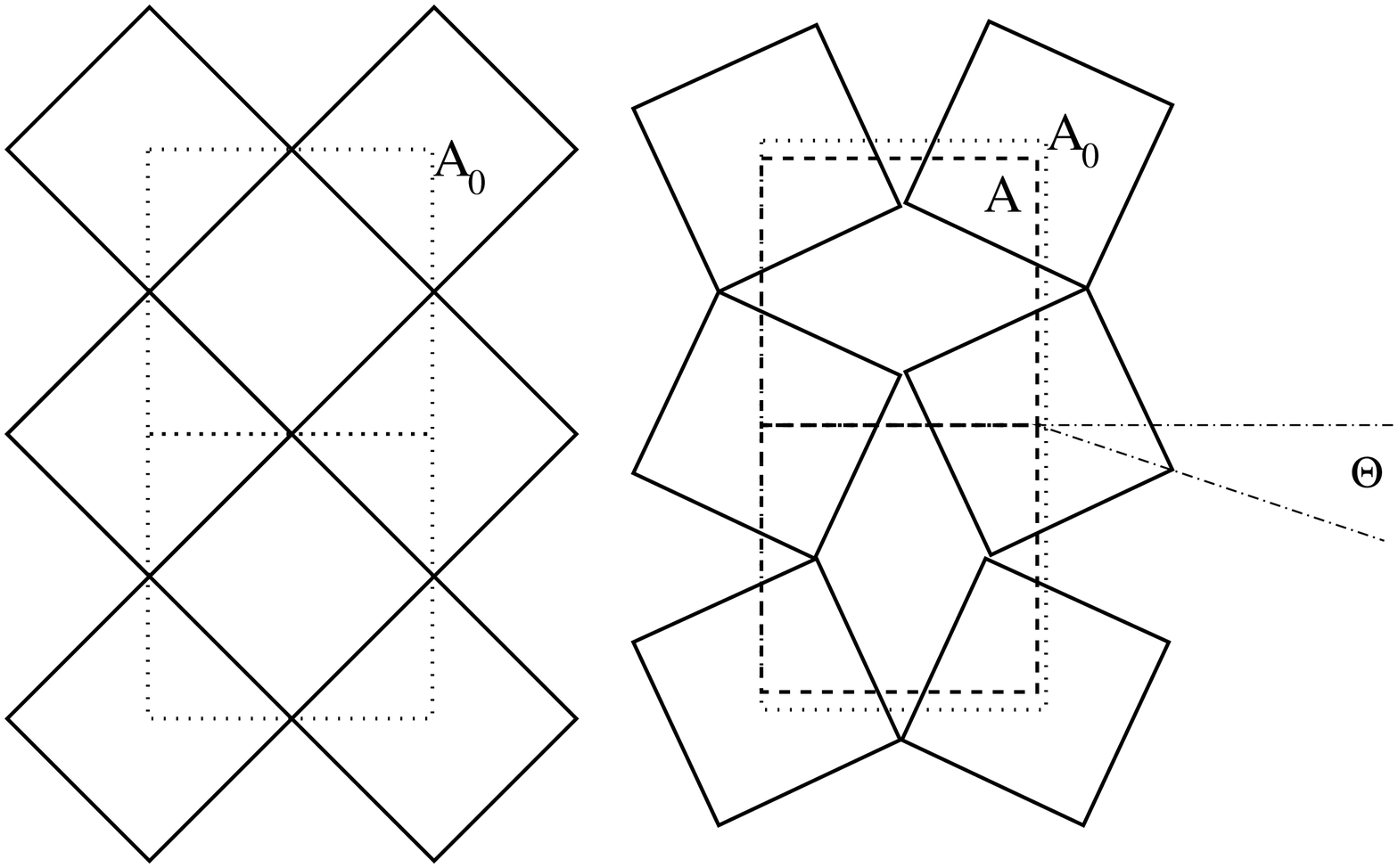,width=150mm,height=100mm}}
\end{picture}
\caption[]{
Schematic picture of rigid unit modes in two spatial dimensions.
By rotation of the unit cells (squares) by the angle $\Theta$
the volume of the system can be reduced.
Comparison of the system volume $A_0$ with $\Theta=0$ (left side, dotted line)
with the system volume $A$ with $\Theta \neq 0$ (right side, dashed line).
}
\label{rum}
\end{center}
\end{figure}

Surprisingly the c- lattice constant obtained by a classical
simulation decreases slightly with increasing temperature. 
This behaviour- which is present in the experimental
data as well- may thus be explained by a purely classical
argument.  
In the literature the `rigid unit modes'~\cite{rum1,rum2,rum3}
are suggested as being responsible for the negative thermal
expansion coefficient. For the sake of a more complete discussion
of our results, we recall this argument here: The idea is that in a covalently
bonded crystal structure one can identify units which are essentially rigid,
because the length of the covalent bonds are essentially rigid, but that
at certain atomic positions where the rigid units are linked together,
they may rotate relative to each other.
These modes may be visualised schematically
as in figure~\ref{rum}  for two dimensional structures.
The quadratic bases in this example may correspond to
SiO$_4$ tetrahedron in $\beta$- quartz. If the angle stabilising
forces in such units are significantly stronger than the forces
at the common O- atoms connecting such tetrahedron, thermal excitation
leads to a rotation of the complete bases.
In contrast to the positive thermal expansion
coefficient caused by the anharmonicities of the interatomic
interactions  such collective modes may cause a negative
thermal expansion coefficient.
The possible mechanism is briefly sketched:
By tilting of the unit cell by the angle $\Theta$
the system volume $A = A_0$ (see Fig.~\ref{rum}) changes to $A = A_0 \cos^2(\Theta)$.
For small angles and harmonic rotational oscillations 
with frequency $\omega$ of the units (i.e. tetrahedra) 
with moment of inertia $I$ the resulting thermal
average of $A$ is~\cite{rum1,rum2,rum3}:
\begin{equation}
A(T) = <A>_T = A_0 (1-<\Theta^2>_T) = A_0 (1- \frac{k_B T}{I \omega^2})
\label{rum6}
\end{equation}
The thermal average of the tilt angle $<\Theta^2>_T$
increases with temperature and thus leads to
a decrease of system volume.
At low temperatures this effect may more than
compensate the `normal' thermal expansion
caused by the motion of single atoms such that
the thermal expansion coefficient is negative.

\subsubsection{Structural details $\beta$-quartz} 
\begin{figure}[hbtp]
\begin{picture}(000,150)
\put(0,0) {\epsfxsize=140mm \epsfbox{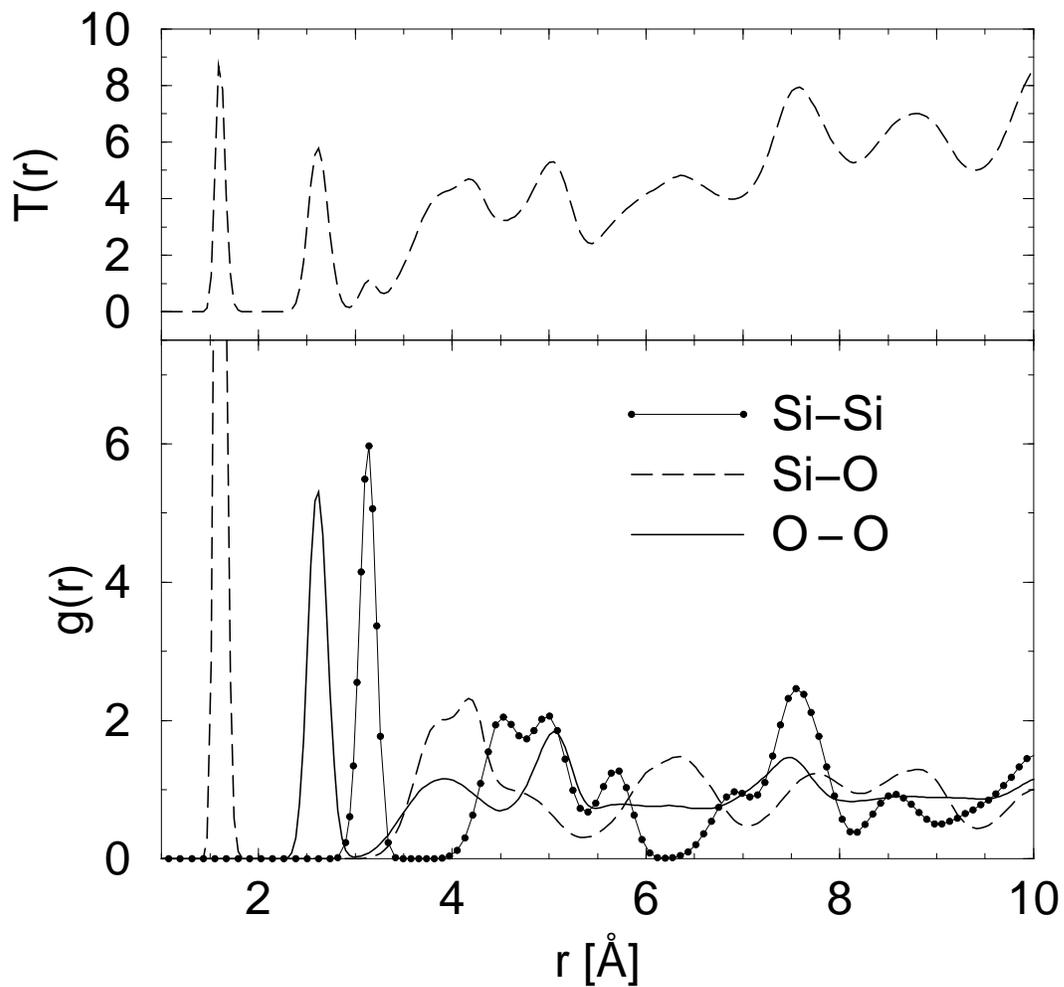}}
\end{picture}
\caption{a) $T(r)$ for $\beta$-quartz
as a function of distance $r$ at 900~K.
b) Corresponding radial distribution function $g(r)$ for Si-Si, Si-O, and O-O
bonds. The curves are normalized such that $g(r) \to 1$ for $r\to\infty$.
}
\label{martin_quartz}
\end{figure}

In order to get more information on the molecular structure, the radial
distribution functions $g(r)$ were computed for $\beta$-quartz,  again
up to a distance of $10\,\AA$. The correlation function $T(r)$, which
follows from the various $g(r)$'s, is shown in Fig.~(\ref{martin_quartz}a)
together with $g_{\rm Si\,Si}$, $g_{\rm Si\,O}$, and  $g_{\rm O\,O}$
in Fig.~(\ref{martin_quartz}b).

It is instructive to compare Fig.~(\ref{martin_quartz}) for $\beta$-quartz
with the corresponding figure for $\beta$-cristobalite,
Fig.~(\ref{martin_crist}). The location of the
peaks in $g(r)$ are nearly identical for nearest neighbor Si-Si, Si-O, and
O-O peaks, only the peaks are broadened in the case of $\beta$-quartz, which
can entirely be attributed to the larger temperature of the $\beta$-quartz
simulation. It can be concluded that the tetrahedra formed by SiO$_4$ units
are basically identical for both lattice structures.
While the radial distribution
functions differ markedly between the two lattices for distances
$r > 3.5\,\AA$, there is a much less pronounced, yet noticeable, effect in
$T(r)$. E.g., $T(r)$ shows a shoulder for $\beta$-quartz at
$r \approx 3.7\,\AA$ which is absent in the simulation of $\beta$-cristobalite.

\subsection{``Size'' of the quantum particles}

\begin{figure}
\begin{center}
\begin{picture}(150,100)
\put(0,0){\psfig{figure=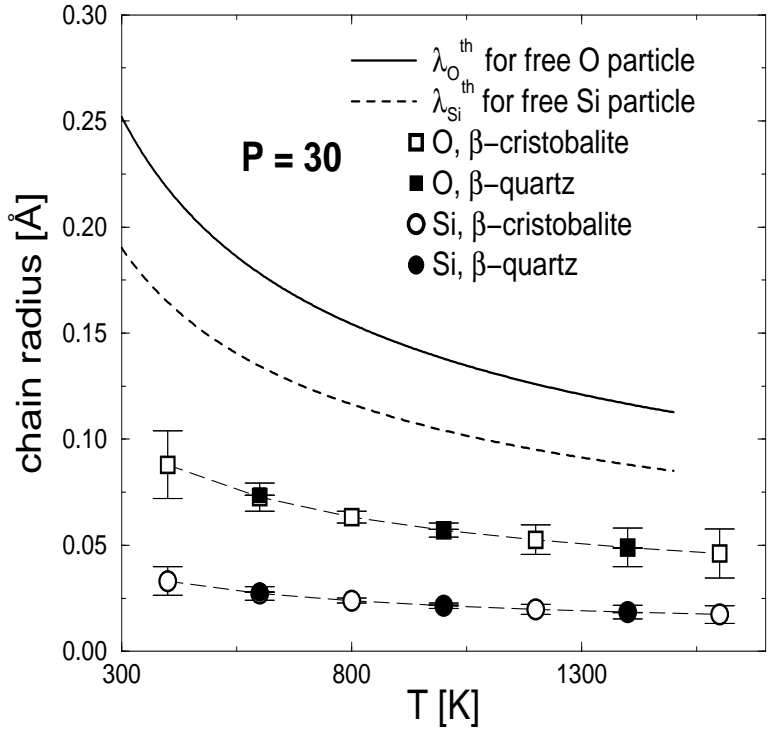,width=100mm,height=100mm}}
\end{picture}
\caption[]{
Quantum chain radii in $\beta$- cristobalite and $\beta$- quartz.
Comparison with the de- Broglie wave lengths of free particles
}
\label{c.sio2}
\end{center}
\end{figure}

By determining the average radius of the quantum-particles 
across the quantum chain their ``size'' can be
mapped as a function of temperature.
In figure~\ref{c.sio2} we compare the simulation results
for these radii in the various lattice structures
with the thermal de-Broglie wave lengths for free O- and Si- atoms.
Apparently this quantity does not depend much on
the particular lattice structure. This may be due to
the fact that the quantum chain radius is determined
by the close surrounding of the particles and the potentials.
In both crystalline structures studied here the basic unit
is a SiO$_4$ tetrahedron and the BKS potential is used.
The de-Broglie wave lengths for the free particles
are much larger than the radii of the particles of the 
crystals. This may be due to the stronger binding and
localisation of the latter.

\section{Summary}

In this paper we present the results of our  PIMC simulations
for crystalline silicates.
With PIMC simulations we are able to analyse the 
thermal properties of crystals at temperatures at which quantum effects
have to be considered as well as anharmonic effects.
In general the agreement with experimental data is better
compared to classical computations. 
For $\beta$--cristobalite it turns out that even at temperatures as
high as 600~K
only with PIMC a good agreement with experimental findings is obtained.
The negative thermal expansion in $\beta$-quartz however may be explained 
by classical effects.

We conclude that for a realistic materials modelling 
the simulation by path integral Monte Carlo techniques
is a useful method which should be used in future studies
of thermal properties of other crystals as well.\\

\acknowledgements
C.R thanks the SCHOTT-- Glaswerke for support and
P.N. thanks the DFG for support (Heisenberg foundation
and SFB~513).
The computations were carried out on the $CRAY\ T3E$ of the
$HLRZ$ at J\"ulich.\\

\section{Appendix A}

In most simulations a rectangular simulation box is used.
It is thus of advantage to map the hexagonal system onto
a larger rectangular box such that the periodic order
results in the hexagonal structure.~\cite{rnb1}

\begin{figure}
\begin{center}
\begin{picture}(150,100)
\put(0,0){\psfig{figure=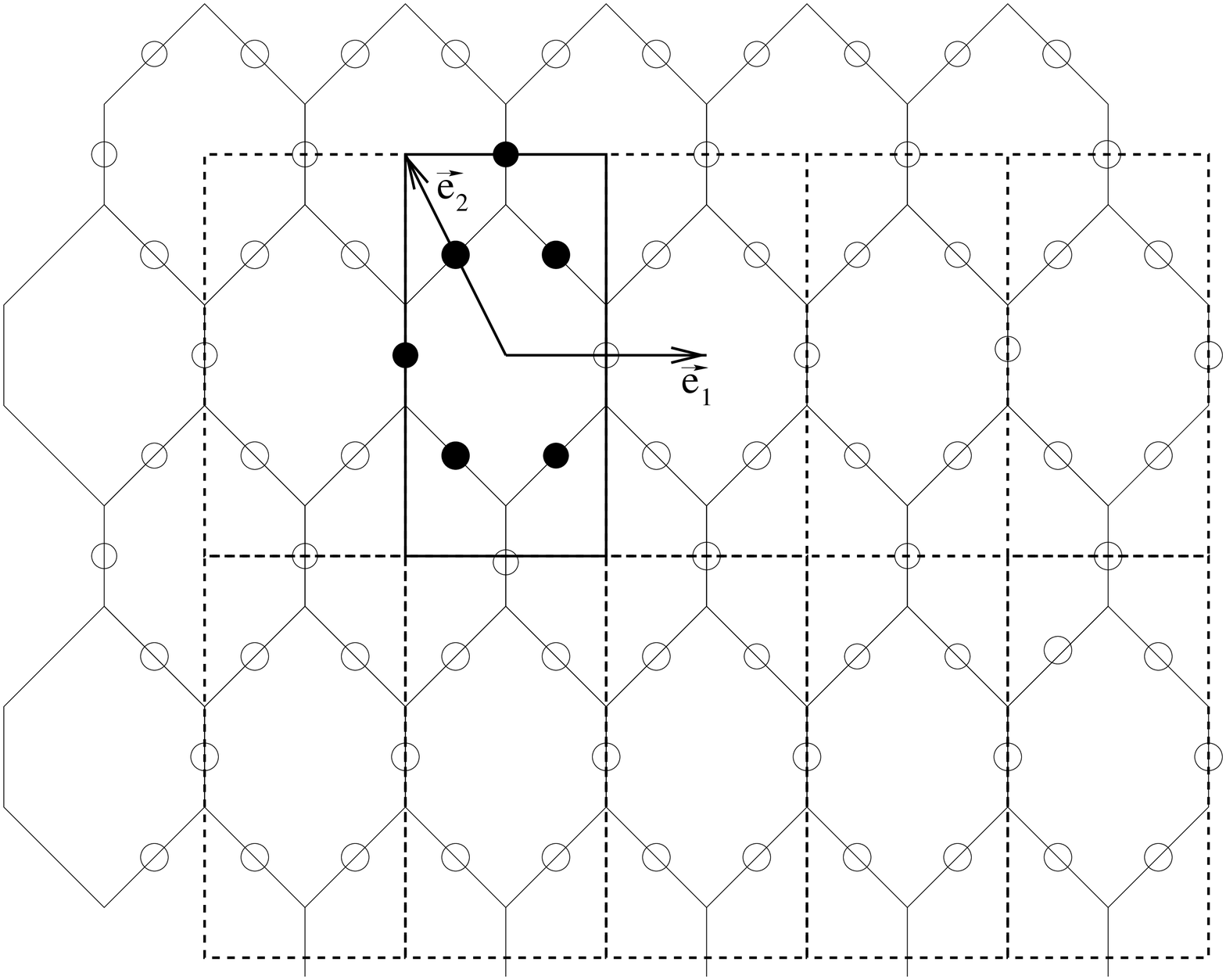,width=150mm,height=100mm}}
\end{picture}
\caption[]{
Hexagonal unit cell of $\beta$-quartz, unit vectors, atom positions and orthogonal
simulation box.
}
\label{hex}
\end{center}
\end{figure}

The lattice structure for $\beta$- quartz is sketched
in Fig.~\ref{hex} and may be described
as follows. In the space group P6$_2$22 the unit
vectors are:
\begin{eqnarray}\nonumber
\vec e_1 &=& (4.9977, 0, 0)\\ \nonumber
\vec e_2 &=& (-2.49885, 4.328135, 0)\\ \nonumber
\vec e_3 &=& (0, 0, 5.4601)\\
\end{eqnarray}
Thus the lattice constants are:
\begin{eqnarray}\nonumber
a & = & 4.9977 \AA \\ \nonumber
b & = & \sqrt{3} a\\ \nonumber
c & = & 5.4601 \AA\\
\end{eqnarray}
The coordinates of the particles in a unit cell
are given in table~III, the Cartesian coordinates in table~IV.

Due to the differences in box lengths one has to reduce
the effect of different surface areas on the simulation results.
This is done by matching sufficiently many unit cells
such that the simulation box is approximately cubic.



\newpage

\newpage

\newpage

\newpage

\newpage


\newpage

\newpage

\newpage

\newpage

\newpage

\newpage
\begin{table}
\label{t1}
\begin{center}
\begin{tabular}{|c||c|c|c|}
\hline
ion pair & A/eV & B/$\AA^{-1}$ & C/eV$\AA^6$ \\
\hline
Si--Si & 872.4  & 15.22 & 23.265\\
O--O  & 1756.98 & 2.846 & 214.75  \\
Si--O & 10722.23 & 4.796 & 70.739 \\
\hline
\end{tabular}
\caption[]{
Parameters in Eq.(\ref{ttam-formel}) for the TTAM-potential. 
}
\end{center}
\end{table}

\begin{table}
\label{t2}
\begin{center}
\begin{tabular}{|c||c|c|c|}
\hline
ion pair & A/eV & B/$\AA^{-1}$ & C/eV$\AA^6$ \\
\hline
Si--Si & 0 & - & 0 \\
O--O & 1388.7730 & 2.76000 & 175.0000 \\
Si--O & 18003.7572 & 4.87318 & 133.5381 \\
\hline
\end{tabular}
\caption[]{
Parameters in Eq.(\ref{ttam-formel}) for the BKS-potential. 
}
\end{center}
\end{table}

\begin{table}
\begin{center}
\begin{tabular}{|c||c|c|c|}
\hline
atom `& x & y & z \\
\hline
Si 1 & 1/2 & 0 & 0 \\
Si 2 & 0 & 1/2 & 2/3 \\
Si 3 & 1/2 & 1/2 & 1/3 \\
O 1 & x & 2x & 1/2 \\
O 2 & \=x & \={2x} & 1/2 \\
O 3 & \={2x} & \=x & 1/6 \\
O 4 & x & \=x & 5/6 \\
O 5 & \=x & x & 5/6 \\
O 6 & 2x & x & 1/6 \\
\hline
\end{tabular}
\caption[]{
Coordinates of the particles in the unit cell of $\beta$-quartz
(see Fig.~10). 
}
\end{center}
\end{table}

\begin{table}
\begin{center}
\begin{tabular}{|c||c|c|c|}\hline atom & x/$\AA$ & y/$\AA$ & z/$\AA$ \\
\hline Si 1 &       2.498850 &    0 &    0 \\Si 2 &      -1.249425 &   2.164068 &   3.640067\\
Si 3 &       1.249425 &   2.164068 &   1.820033\\O 1  &       0  &  1.793579  &  2.730050\\
O 2  &       2.498850 &   2.534556 &   2.730050\\
O 3  &       0.945565 &   3.431346 &    0.910017\\
O 4  &       -.945565 &   3.431346 &   4.550083\\
O 5  &       3.444415 &    0.896790 &   4.550083\\
O 6  &       1.553285 &    0.896790 &    0.910017\\
\hline
\end{tabular}
\caption[]{
Cartesian coordinates of the particles in the unit cell of $\beta$-quartz. 
}
\end{center}
\end{table}


\begin{references}


\bibitem{rlmb}
G.C Rutledge, D.J. Lacks, R. Martonak, K. Binder;
J. Chem. Phys. {\bf 108}, 10274 (1998).

\bibitem{rnb3} R.P. Feynman and A.R. Hibbs,
{\it Quantum Mechanics and Path Integrals} (McGraw--Hill,
New York, 1965); H.F. Trotter, Proc. Am. Math. Soc.
{\bf 10} , 545 (1959); Commun. Math. Phys. {\bf 51}, 183 (1976);
J.A. Barker, J. Chem. Phys. {\bf 70}, 2914 (1979);
K. S. Schweizer, R. M. Stratt, D. Chandler, P. G. Wolynes,
J. Chem. Phys.  {\bf 75}, 1347 (1981);
D. M. Ceperley, Rev. Mod. Phys. {\bf 67}, 279 (1995);
P. Nielaba,
in {\it Annual Reviews of Computational Physics~V}
edited by D. Stauffer (World Scientific, Singapore), pp.~137 (1997).

\bibitem{rnb10} K. Vollmayr, Ph.D thesis, Mainz (1995).

\bibitem{tsuneyuki88} S. Tsuneyuki, M. Tsukada, H. Aoki, Y. Matsui;
Phys. Rev. Lett. {\bf 61}, 869 (1988).

\bibitem{bks} B. van Beest, G. Kramer, R. van Santen;
Phys. Rev. Lett. {\bf 64}, 1955 (1990).

\bibitem{schroeder96} K.-P. Schr\"oder, J. Sauer;
J. Phys. Chem. {\bf 100}, 11043 (1996).

\bibitem{frenkel96}
D. Frenkel and B. Smit,
{\it Understanding Molecular Simulation: From Algorithms to Applications},
(Academic Press, San Diego, 1996).

\bibitem{kathrin} K. Vollmayr, W. Kob, K. Binder;
Phys. Rev. {\bf B 54}, 15808 (1996).

\bibitem{rnb1} Chr. Rickwardt, Ph.D thesis, Mainz (1998).

\bibitem{rnb8} I.P. Swainson, M.T. Dove, Phys. Chem.
 Minerals {\bf 22}, 61 (1995). 

\bibitem{wyckoffa} R. Wyckoff; Am. J. Sci {\bf 9}, 448 (1925).

\bibitem{wyckoffb} R. Wyckoff; Z. f. Krist. {\bf 62}, 189 (1925).

\bibitem{wright75} A. Wright, A. Leadbetter; Phil. Mag. {\bf 31},
 1391 (1975).

\bibitem{hatch91} D. Hatch, S. Ghose; Phys. Chem. Min. {\bf 17}, 
554 (1991).

\bibitem{dove92a} M. Dove, A. Giddy, V. Heine;
 Ferroelectrics {\bf 136}, 33 (1992).

\bibitem{dove92b} M. Dove, A. Giddy, V. Heine; 
Trans. Am. Cryst. Assoc. {\bf 27}, 697 (1992).

\bibitem{giddy93} A. Giddy, M. Dove, G. Pawley, V. Heine;
 Acta Cryst. {\bf A 49}, 697 (1993).

\bibitem{dove97}
M. T. Dove, D. A. Kreen, A. C. Hannon, and I. P. Swainson,
Phys. Chem. Minerals {\bf 24}, 311 (1997). 

\bibitem{swainson95b}
I. P. Swainson and M. T. Dove,
J. Phys. Cond.Matt. {\bf 7} 1771 (1995).



\bibitem{tse91} 
J. Tse, and D. Klug;
 Phys. Rev. Lett. {\bf 67}, 3559 (1991);
Science {\bf 255}, 1559 (1992).

\bibitem{tse92a} J. Tse, D. Klug, Y. LePage;
 Phys. Rev. Lett. {\bf 69}, 3647 (1992).

\bibitem{tse92b} J. Tse, D. Klug, Y. LePage;
  Phys. Rev. {\bf B 46}, 5933 (1992).

\bibitem{liu93} F. Liu, S. H. Garofalini, R. D. King--Smith, D. Vanderbilt;
 Phys. Rev. Lett. {\bf 70}, 2750 (1993).

\bibitem{axe70} J. Axe, G. Shirane;
 Phys. Rev. {\bf B 1}, 342 (1970).

\bibitem{kihara90} K. Kihara;
 Eur. J. Mineral. {\bf 2}, 63 (1990).

\bibitem{rum1} P. Welche, V. Heine, M. Dove;
 Phys. and Chem. of Minerals. Berlin: Springer (1997).

\bibitem{rum2} M. Dove, V. Heine, K. Hammonds;
 Mineral. Mag {\bf 59}, 629 (1997).

\bibitem{rum3} M. Gambhir, M.T. Dove, V. Heine;
 Phys. Chem. Minerals {\bf 26}, 484 (1999).


\end{references}
\end{document}